\documentclass[a4paper,11pt]{article}
\usepackage{pos}
\usepackage{subcaption}

\title{Cosmic Ray Acceleration by Turbulence-Driven Magnetic Reconnection and the Origin of the Neutrinos in NGC 1068}
\ShortTitle{CR Acceleration by Turbulence-Driven Magnetic Reconnection in NGC 1068}

\author*[a]{Luana Passos-Reis}
\author[a]{Elisabete M. de Gouveia Dal Pino}
\author[b]{Juan Carlos Rodríguez-Ramírez}
\author[a]{Giovani H. Vicentin}

\affiliation[a]{Instituto de Astronomia, Geof\'{i}sica e Ci\^{e}ncias Atmosf\'{e}ricas (IAG-USP), Universidade de S\~{a}o Paulo, \\
1226 Rua do Mat\~{a}o, CEP: 05508-090, S\~{a}o Paulo - SP, Brazil.}

\affiliation[b]{Centro Brasileiro de Pesquisas Físicas (CBPF), \\
150 Rua Dr. Xavier Sigaud, CEP: 22290-180, Rio de Janeiro - RJ, Brazil.}

\emailAdd{$^{*}$luana.passos.reis@usp.br}

\abstract{The Seyfert Type II galaxy NGC 1068 has been identified as a potential neutrino source by IceCube, with a 4.2$\sigma $ significance detection of a 79$^{+22}_{-20}$ neutrino excess from 2011 to 2020 \cite{IceCube_2022}, despite the absence of a gamma-ray counterpart. The observed high-energy neutrino emission indicates the presence of a hadronic component, along with strong gamma-ray absorption, likely via pair production, and efficient particle acceleration.  
In this work, we investigate turbulence-driven magnetic reconnection as a mechanism for particle acceleration in the coronal accretion flow surrounding the central black hole. We develop a one-zone model for both acceleration and emission, following the framework of \cite{dalpino_lazarian_2005} and \cite{kadowaki_etal_15} to explore how fast magnetic reconnection in the inner coronal disk region accelerates protons and electrons, shaping the spectral energy distribution (SED).  
Our model incorporates strong pair production attenuation and interactions with optical, ultraviolet (OUV), and X-ray photon fields in the corona, which serve as effective targets for proton interactions. Unlike recent studies, we find that particle acceleration to the extreme energies required to explain observations is primarily driven by first-order Fermi acceleration within the turbulent reconnection layers in a large scale current sheet, rather  than by drift acceleration. 
Additionally, we demonstrate that accelerated protons primarily lose energy through photopion interactions with the OUV background, subject to important constraints from the coronal X-ray emission.
}

\ConferenceLogo{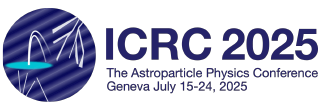}

\FullConference{39th International Cosmic Ray Conference (ICRC2025)\\
 15–24 July 2025\\
Geneva, Switzerland\\}

\begin{document}
\maketitle

\section{Introduction}

The nearby active galaxy NGC1068, also known as Messier 77, has been on spotlight as a multi-messenger source, since the IceCube Collaboration report about a significant excess of high-energy neutrinos associated to this source, with $79_{-20}^{+22}$ events of neutrino excess at a $4.2 \sigma $ significance. The data taken from 2011 to 2020 by the IceCube Collaboration \citep{IceCube_2022} provides a compelling evidence for a powerful hadronic component. However, this detection is notable for the absence of a corresponding TeV gamma-ray counterpart, which are often produced within the same mechanisms as the high-energy neutrinos. This absence suggests that the neutrinos may originate from a highly obscured environment, such as the 
the corona-accretion disk region \citep{Murase2022}, where gamma-rays are efficiently absorbed by photon fields, likely via pair production ($\gamma \gamma $ interactions due to the optical depth).

This very high-energy (VHE) neutrino excess presents a fundamental challenge: which physical mechanism(s) would be able to efficiently accelerate protons up to the required extreme energies, in the inner magnetized regions of the accretion flow?
While some recent works have explored particle acceleration by plasmoid-driven magnetic reconnection, 
their findings are elusive, 
often requiring additional assumptions about particle pre-acceleration or complex geometries, envolving several combined processes and different acceleration/emission regions for the neutrinos to be produced 
\citep[e.g.][and references therein]{Karavola2025,Fiorillo2025}. 
Moreover, recent high-resolution MHD simulations suggest that plasmoid-driven reconnection is slow and may not be the dominant 
acceleration mechanism at these environments \citep{Vicentin_etal_2025}.

In this work, we investigate a different particle acceleration scenario: turbulence-driven magnetic reconnection in the accretion disk-corona. Our model, based on previous work \citep{dalpino_lazarian_2005, dalpino_piovezan_2010, kadowaki_etal_15}, assumes that particle acceleration is governed predominantly by  first-order Fermi process within the turbulent reconnection layers \citep{dalpino_lazarian_2005} (see also de Gouveia Dal Pino et al., these Procs., for a review on this mechanism).  Unlike other models where reconnection acceleration grows linearly with time (e.g., via drift or escape time constraint), our approach allows for a process where the acceleration rate is independent of the particle's energy (as expected in the Fermi process),
making it a highly efficient and robust mechanism.

In the following, we present preliminary results of this model, which will be detailed in forthcoming work.


\section{The Coronal Model and Particle Acceleration}

\begin{figure}[h!]
    \centering
    \includegraphics[width=0.5\textwidth]{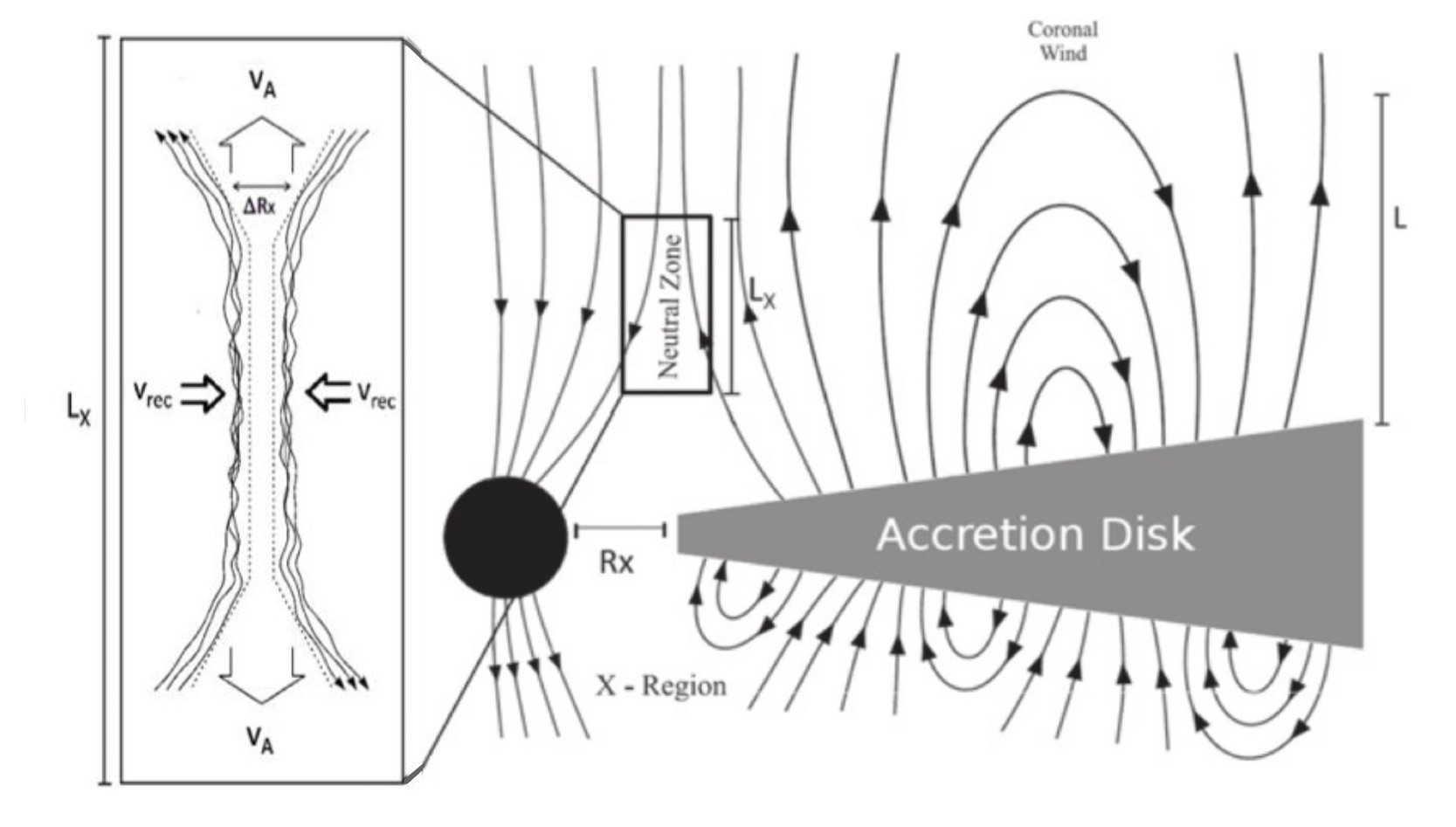}
    \caption{Schematic representation of  our model. Field lines tied to the BH horizon meet oppositely oriented disk lines in the corona, forming a large-scale current sheet embedded in turbulence. 
    The main features include the inner disk radius, $R_X$, and the turbulent reconnection region, characterized by the current sheet (neutral zone) width, $\Delta R_X$, and height of the reconnection region, $L_X$. Particles are accelerated and emit within this region. $L_X$ is constrained by the coronal extension 
    $L$ (see text for more details). 
    Adapted from  \citep{dalpino_lazarian_2005, kadowaki_etal_15}.}
    \label{fig:model_image}
\end{figure}

We model the central engine of NGC 1068 as a supermassive black hole surrounded by a geometrically thin, optically thick accretion disk \citep{shakura_sunyaev_73}, 
and a hot, magnetized corona, following the one-zone model in \citep{dalpino_lazarian_2005, dalpino_piovezan_2010, kadowaki_etal_15}, as shown in Figure \ref{fig:model_image}.
Magnetic field lines anchored in the black hole horizon interact with oppositely directed lines from the accretion disk in the corona, giving rise to a large-scale current sheet.
Differential rotation and instabilities such as the magnetorotational instability (MRI) and Parker-Rayleigh-Taylor instability drive turbulence, leading to fast magnetic reconnection (\citep{Lazarian_Vishniac_1999, Kadowaki_etal_2018}) in this zone, marked in the image as the Neutral Zone. This process releases  magnetic power, a fraction of which may be channeled into accelerating particles, that will then cascade and generate the observed emission.

In our model, the physical conditions in the acceleration  region at the corona are described by a set of free and model-dependent  parameters.
The model-dependent parameters are derived in 
\citep{dalpino_lazarian_2005, dalpino_piovezan_2010, kadowaki_etal_15}, and are summarized below:
\begin{align}
\label{eq:full_model_equations}
\left\{
\begin{aligned}
B_{c} &\simeq 9.96 \times 10^{8} r_{X}^{-5/4} \dot{m}^{1/2} m^{-1/2} \text{ G} \\
\dot{W_{B}} &= 1.66 \times 10^{35} \Gamma^{-1/2} r_{X}^{-5/8} l^{-1/4} l_{X} q^{-2} \dot{m}^{3/4} m \\
n_{c} &\simeq 8.02 \times 10^{18} \Gamma^{1/2} r_{X}^{-3/8} l^{-3/4} q^{-2} \dot{m}^{1/4} m^{-1} \text{ cm}^{-3} \\
\Delta R_{X} &\simeq 11.6 \Gamma^{-5/4} r_{X}^{31/16} l^{-5/8} l_{X} q^{-3} \dot{m}^{-5/8} m \text{ cm} \\
T_{c} &\simeq 2.73 \times 10^{9} \Gamma^{1/4} r_{X}^{-3/16} l^{1/8} q^{-1} \dot{m}^{1/8} \text{ K} \\
T_{d} &\simeq 3.71 \times 10^{7} \alpha^{-0.25} r_{X}^{-0.37} m^{0.25} \text{ K}
\end{aligned}
\right.
\end{align}
Where $T_d$  is the disk temperature, and $B_{c}$, $T_c$, and $n_c$ are the coronal magnetic field, temperature and number density, respectively,  at the reconnection coronal region. These  quantities depend on the following  dimensionless  parameters: 
the black hole mass, $m$, expressed in units of solar mass ($M_{\odot}$), the accretion rate, $\dot{m}$, expressed in Eddington accretion units ($\dot{M}_{\rm Edd}$),
the distance from the  BH to the reconnection site $r_X = R_X / R_{\rm Sch}$,  the turbulent reconnection region height $l_X = L_X / R_{\rm Sch}$, and the extension of the corona $l = L / R_{\rm Sch}$, expressed in terms of $R_{\rm Sch}$,
the Schwarzschild radius.
The factors, $\Gamma $ and $q$,  account for special-relativistic effects on the Alfvén speed and geometry correction of the inner disk, respectively, and are given by
$\Gamma = \left(1 + \left( \frac{v_{A0}}{c} \right)^{2} \right)^{-1/2}$, 
$q = \left[ 1 - \left(\frac{3 R_{\rm Sch}}{R_{X}} \right)^{1/2} \right]^{1/4}$, 
where  $v_{A0} = \frac{B}{\sqrt{4 \pi \rho}}$.  
The magnetic recconection power, $\dot{W_{B}}$, which is released by turbulence-driven fast reconnection in the surrounds of the BH is essential for characterizing the efficiency of particle acceleration.
The model uses 
the proton rest mass ($m_H$) and a mean molecular weight of $\mu \sim 0.6$, so that $\rho_c = \mu m_{H} n_{c}$.


\begin{table}[htbp]
    \centering
    \caption{Observational Parameters for NGC 1068.}
    \begin{tabular}{ccc}
    \hline
    \hline
    Parameter & Value & Unit \\
    \hline 
        $m$ (Black Hole Mass) & $2 \times 10^7$ & [$M/M_\odot$] \\
        $\dot{m}$ (Accretion Rate) & 0.55 & [$\dot{M}/\dot{M}_{\rm Edd}$] \\
        $d$ (Distance) & $10.1$ & [Mpc] \\
        $z$ (Redshift) & $0.00379$ & \\
    \hline
    \end{tabular}
    \label{obs_table}
\end{table}





The free parameters of the model are given in Table \ref{obs_table}, and the model-dependent ones in Table \ref{table_parameters}, for a given  set of fiducial values of  $r_x$, $l_X$, and $l$. A larger parametric space will be discussed in a forthcoming work. 


\begin{table}[htbp]
    \centering
    \caption{Model-derived parameters, for a given  set of fiducial values of  $r_x$, $l_X$, and $l$, adopted for computing cooling rates and the Spectrum Energy Distribution (SED) of NGC1068.}
    \begin{tabular}{lccc}
    \hline
    \hline
    This Work & Parameter & Value & Unit \\
    \hline
    Coronal Magnetic Flux Tube Height & $l$ & 30 & [$L/R_{\rm Sch}$] \\
    Height of Reconnection Region & $l_X$ & 30 & [$L_X/R_{\rm Sch}$] \\
    Inner radius of disc & $r_X$ & 3.3 & [$R_X/R_{\rm Sch}$] \\
    \\
    \hline
    \hline
    Model-derived Parameters & & \\
    \hline
    Coronal Magnetic Field & $B_c$ & $3.6 \times 10^{4}$ & [G] \\
    Coronal Particle Density & $n_c$ & $5.8 \times 10^{10}$ & [cm$^{-3}$] \\
    Coronal Temperature & $T_c$ & $5.7 \times 10^{9}$ & [K] \\
    Disc Temperature & $T_d$ & $6.2 \times 10^{5}$ & [K] \\
    Width of Current Sheet & $\Delta R_{X}$ & $1.6 \times 10^{11}$ & [cm] \\
    \quad \quad [in Schwarzschild Radii: $R_{\rm Sch}$] & & $\simeq 0.028$ & [$R_{\rm Sch}$] \\
    Reconnection Power Released & $\dot{W}_{B}$ & $6.1 \times 10^{43}$ & [erg\,s$^{-1}$] \\
    \hline
    \hline
    \end{tabular}
    \label{table_parameters}
\end{table}


Our approach, adapted from previous work, allows us to consistently derive the key physical parameters of the corona. It emphasizes first-order Fermi acceleration within the turbulent reconnection sites, as the dominant and highly efficient mechanism. This process leads to energy-independent acceleration time compared to the energy-dependent drift acceleration (see \citep{dalpino_tania_2025}, and references therein),
enabling protons to reach the energies required for neutrino production. 

\section{Particle Interactions and Multi-messenger Emission}

In order to investigate the particle energetics of the system, 
we model two photon fields to estimate photo-hadronic interaction timescales: (i) a blackbody component from the accretion disk with temperature $T_d$ (based on a Shakura-Sunyaev \citep{shakura_sunyaev_73} disk with a viscosity parameter of $\alpha \simeq 0.1$, see Table~\ref{table_parameters}), which serves as a proxy for the optical, ultraviolet (OUV) background, and (ii) an X-ray field constrained by NuSTAR and XMM–Newton observations of NGC~1068 \citep{Bauer2015, Marinucci2016}.

The coronal X-ray emission is modeled using two different approaches to reproduce the observed spectra: one assuming a fraction of the total magnetic power released $\eta \sim 0.7$, for the \citep{Bauer2015} data, and another with a fraction of the accretion power $\sim 0.03$, for the \citep{Marinucci2016} data. Both models utilize a single broken power-law spectrum with a peak energy ($E_0 \sim 200$ eV), a cutoff energy ($E_{\rm cut} \sim 6 \times 10^{4}$–$5 \times 10^{5}$ eV), and a spectral index ($\alpha \sim 2.0$–$2.05$). 
The remaining magnetic power ($\sim 30\%$) is assumed to be injected into the accelerated protons with a power-law index of $1.7$, modeled as a function of the particle number density and the interaction timescale. These protons will undergo subsequent hadronic interactions, primarily proton-proton (p-p), which produce high-energy neutrinos and gamma-rays through the decay of pions ($\pi^{0}$ and $\pi^{\pm }$).




\begin{figure}[h!]
    \centering
    \includegraphics[width=0.55\textwidth]{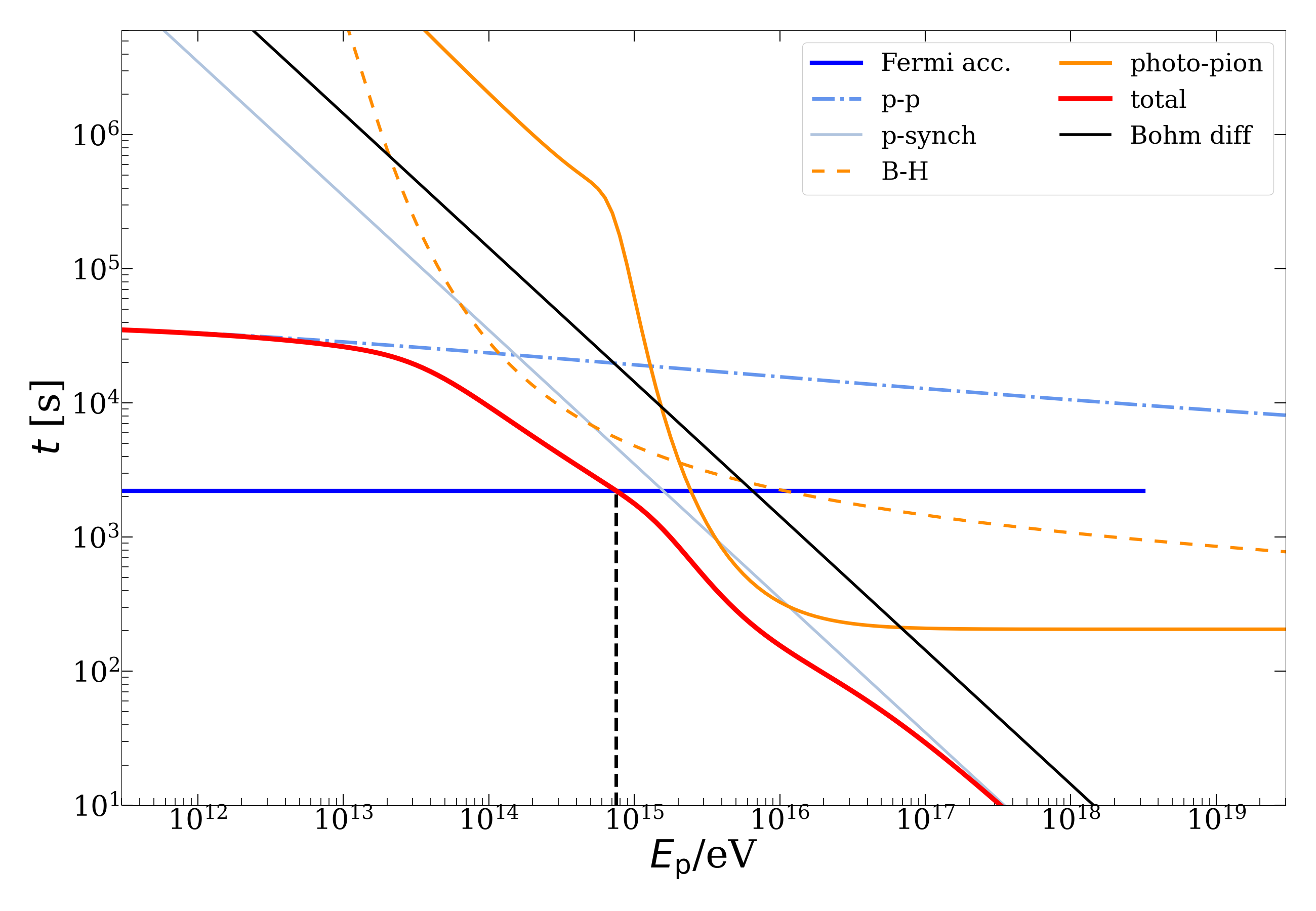}
    \caption{Hadronic acceleration and cooling timescales in the NGC 1068 corona. This figure shows the proton energy-independent acceleration timescale due to first-order Fermi  and the cooling processes whithin the turbulent reconnection layer in the corona. The solid blue curve represents the 
    acceleration timescale, which exhibits a cutoff right after $10^{18}$ eV, corresponding to the energy where the proton's Larmor radius becomes comparable to the reconnection sheet width ($\Delta R_X$). Beyond this energy, drift acceleration would become the dominant mechanism. Energy loss timescales are shown for synchrotron radiation (solid gray), proton-proton (p-p) interactions (light blue, dot-dashed line), and photo-hadronic interactions (photo-pion, p$\gamma $, in solid orange; Bethe-Heitler pair production in dashed orange). The photo-hadronic processes account for both disk blackbody and X-ray photon fields (orange curves). Bohm diffusion (solid black) is included for comparison. The total energy loss timescale is represented by the solid red curve. The intersection of the acceleration and total loss curves determines the maximum proton energy achievable 
    $\simeq 10^{14}$ eV. 
    }
    \label{fig:cool_HAD}
\end{figure}


Our model
focuses on explaining the high-energy neutrino emission
from protons that are accelerated via turbulence-driven magnetic reconnection and subsequently undergo hadronic and photo-hadronic interactions,
cascading and producing the multi-messenger signal. The particle interactions were approximated through timescale estimates, following the first-order Fermi acceleration and cooling mechanisms described in detail in the recent work \cite{dalpino2024}.

Specifically, the acceleration time due to  first-order Fermi  within a turbulent current sheet is given by (\citep{Xu_Lazarian_2022, dalpino_tania_2025}; see also de Gouveia Dal Pino et al., these Procs.):
$t_{\rm acc} \simeq \frac{4 \Delta R_X}{c d}$,
where 
$d \approx \frac{2\beta_{\rm in}(3\beta_{\rm in}^2+3\beta_{\rm in}+1)}{3(\beta_{\rm in}+0.5)(1-\beta_{\rm in}^2)}$,
and $\beta_{\rm in}= v_{\rm rec}/c$, with a reconnection velocity 
$\beta_{\rm in}=v_{\rm rec}/c \simeq 0.01$, 
from numerical simulations. The interplay between these acceleration and cooling mechanisms, shown in Figure \ref{fig:cool_HAD}, defines the maximum achievable proton energy as the intersection of the acceleration and total energy loss curves. Additionaly, we model the $\gamma $-ray annihilation through pair production channels, due to the optical depth opacity of our system, given the two radiation fields mentioned (blackbody radiation from disk, and coronal X-ray constraints).

\begin{figure}[h!]
    \centering
    \begin{subfigure}{0.8\textwidth}
    \centering
    \includegraphics[width=\linewidth]{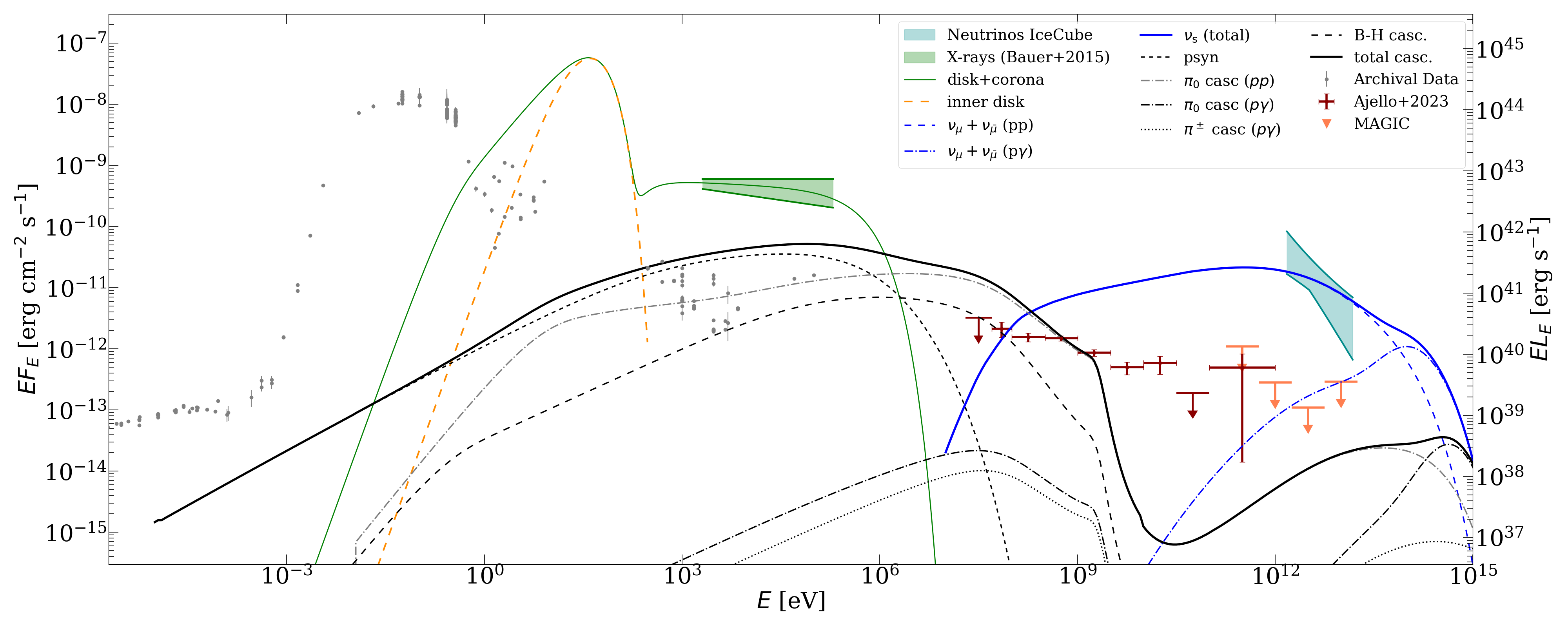}
    \caption{SED constrained by \citep{Bauer2015} X-ray data, representing a typical coronal state. The model shows tension with Fermi-LAT data points in the GeV range.}
    \label{fig:first_sed}
    \end{subfigure}
    \hfill
    \begin{subfigure}{0.8\textwidth}
    \centering
    \includegraphics[width=\linewidth]{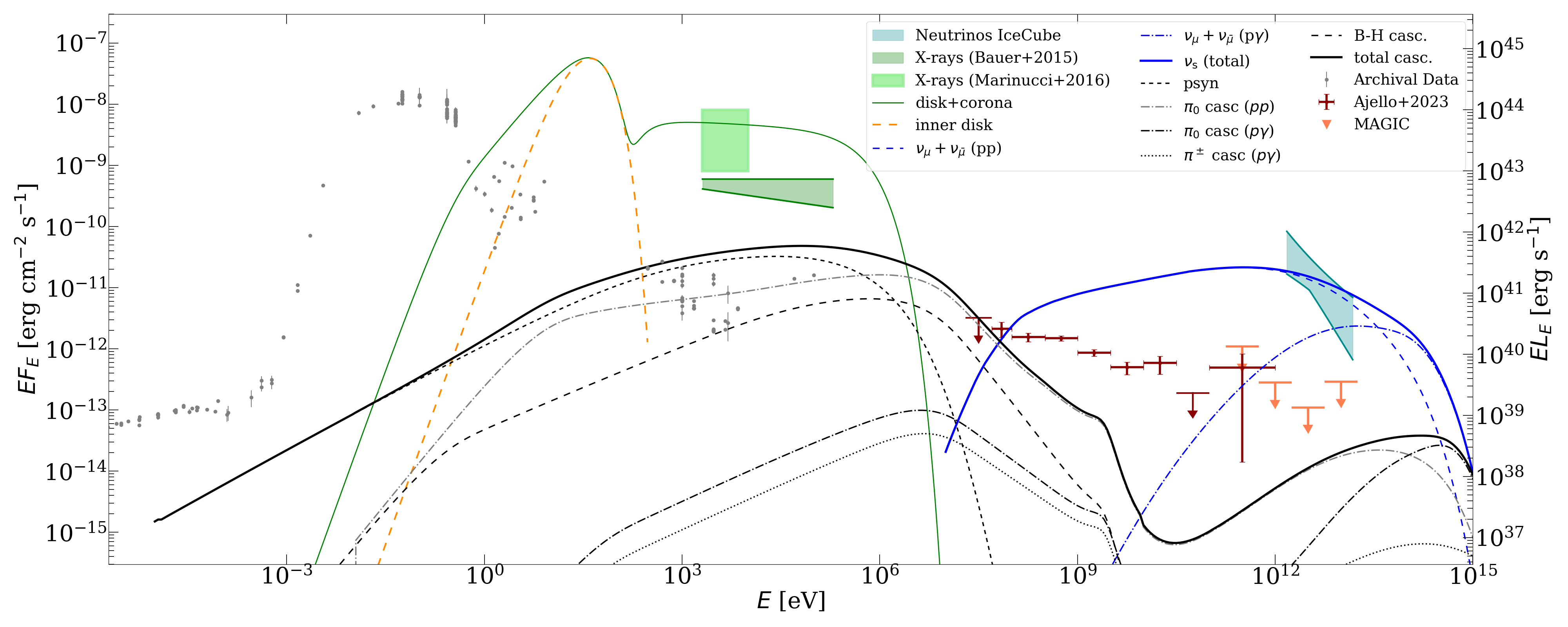}
    \caption{SED constrained by \citep{Marinucci2016} X-ray data from a less obscured period. This model successfully resolves the tension with Fermi-LAT data, highlighting the effects of intrinsic X-ray variability.}
    \label{fig:second_sed}
    \end{subfigure}
    \caption{Two possible Spectral Energy Distributions (SEDs) of the source NGC1068. In both panels it is shown IceCube neutrinos (shaded blue area), MAGIC constraints on $\gamma $-rays (orange upper limits) along with the Archival Data (gray points), from \cite{IceCube_2022}. The dashed orange line represents the modeled blackbody radiation from the accretion disk, while the solid green line shows the combination of the disk and coronal X-ray photon fields. The dark red points show Fermi-LAT detections in the GeV band, presenting variability due to the modeling of X-rays (\citep{Bauer2015} or \citep{Marinucci2016}). The solid black curve represents the total modeled emission from a photo-hadronic cascade (sum of dashed and dotted-dashed cascading lines and physical processes), and the solid blue curve shows the total neutrino emission, with its normalization scaled to represent only the muon neutrinos. The model does not account for Extragalactic Background Light (EBL) gamma-ray attenuation.}
    \label{fig:combined_seds}
\end{figure}

The Spectral Energy Distribution (SED) (Figure~\ref{fig:combined_seds}) provides a comprehensive view of the high-energy processes in NGC 1068's corona. The SED is presented in two panels to illustrate the effect of X-ray variability. The upper panel (Figure \ref{fig:first_sed}) uses the \citep{Bauer2015} X-ray data, representing an average spectral state. This model shows some tension with the Fermi-LAT data in the GeV range, suggesting that other physical processes may contribute to the emission in this region. The lower panel (Figure \ref{fig:second_sed}), however, uses the \cite{Marinucci2016} X-ray data, obtained during a less obscured observation period. This dataset provides tighter constraints on the coronal emission and successfully resolves the tension with the Fermi-LAT data, highlighting the source's intrinsic variability and demonstrating a consistent fit across the spectrum.

An important aspect of our model is the efficient absorption of gamma-rays through photon-photon annihilation ($\gamma \gamma $ annihilation), along with the combined blackbody and X-ray photon fields. This process suppresses the gamma-ray emission, keeping our model below the MAGIC upper limits while still successfully reaching energies sufficient to explain the observed neutrino flux in Figure~\ref{fig:combined_seds}. This consistency validates our approach.

\section{Summary and Discussion}

In this work, we present a model where the high-energy particle emission in the NGC1068 corona is driven by turbulence-driven magnetic reconnection particle acceleration.  
Our model reproduces the observed neutrino signal from NGC 1068, while remaining consistent with gamma-ray observational upper limits. The maximum energy of accelerated protons is determined by the interplay between a first-order Fermi acceleration mechanism and the total cooling of radiative processes. The calculated SED, presented in Figure \ref{fig:combined_seds}, based on the hadronic interactions shown in Figure \ref{fig:cool_HAD}, shows that our model is able to accelerate protons up to energies larger than $10^{14}$ eV, that would subsequently cascade and produce neutrinos, reaching the matching energy for the IceCube data from 2011 until 2020 \cite{IceCube_2022}. We present two possible SED modeling 
with different X-ray coronal descriptions constrained by Chandra NuSTAR spectroscopy observations.

This analysis helps to mitigate the tension with the initial Fermi-LAT upper limit, highlighting the source's intrinsic variability, illustrating the impact of this specific X-ray luminosity data on the overall SED shape, further demonstrating the complex interplay between different energy bands. These tests were conducted to explore the influence of X-ray variability but are not central to the primary findings of this work. Another key result is the strong suppression of gamma-ray emission due to efficient photon-photon annihilation, which keeps our model below the constraints set by MAGIC upper limits, in the high energy range, thus explaining the lack of a gamma-ray counterpart. This result highlights the consistency of our model and suggests that turbulent-driven magnetic reconnection in the corona could be a viable mechanism for particle acceleration and for powering multi-messenger emission in obscured active galactic nuclei. The observed gamma-ray upper limits at TeV energies by MAGIC,
in Figure \ref{fig:combined_seds}, are possibly produced at the AGN outflow region (e.g., \citep{Peretti_2023}).

A final remark is warranted. The preliminary results presented here,  do not include general relativistic curvature effects, which may become relevant for particle and photon trajectories at distances $r_X \lesssim 6 R_{\rm sch}$ \citep{kadowaki_etal_15}. These effects will be incorporated in forthcoming work in preparation.

\end{document}